\documentclass[sigconf]{acmart}
\AtBeginDocument{%
  }

\setcopyright{acmlicensed}
\copyrightyear{2025}
\acmYear{2025}
\acmDOI{3736229.3736269}
\acmConference[PW '25]{Provenance Week @ SIGMOD}{June 27, 2025}{Berlin, Germany}
\acmISBN{978-1-4503-XXXX-X/2018/06}

\usepackage{listings}

\begin{document}

\title{Fast Capture of Cell-Level Provenance in Numpy}

\author{Jinjin Zhao}
\email{j2zhao@uchicago.edu}
\affiliation{%
  \institution{University of Chicago}
  \city{Chicago}
  \state{IL}
  \country{USA}
}

\author{Sanjay Krishnan}
\email{skr@uchicago.edu}
\affiliation{%
  \institution{University of Chicago}
  \city{Chicago}
  \state{IL}
  \country{USA}
}


\begin{abstract}
Effective provenance tracking enhances reproducibility, governance, and data quality in array workflows. However, significant challenges arise in capturing this provenance, including: (1) rapidly evolving APIs, (2) diverse operation types, and (3) large-scale datasets. To address these challenges, this paper presents a prototype annotation system designed for arrays, which captures cell-level provenance specifically within the \texttt{numpy} library. With this prototype, we explore straightforward memory optimizations that substantially reduce annotation latency. We envision this provenance capture approach for arrays as part of a broader governance system for tracking for structured data workflows and diverse data science applications.
\end{abstract}



\maketitle

\section{Introduction}
Array-structured programs are ubiquitous in data science tools, such as \texttt{numpy} arrays \cite{numpyarray}, \texttt{pandas} dataframes \cite{pandasdataframe}, \texttt{pytorch} tensors \cite{pytorch}, spreadsheets, and similar frameworks. Capturing cell-level provenance from array operations can significantly improve the governance, reproducibility, and overall quality of data science pipelines. However, existing data systems still face significant challenges in automatically capturing cell-level provenance.

We have identified three key characteristics of modern data pipelines that contribute to these challenges: (1) Changing APIs, (2) Diverse Operations, and (3) Scale of Datasets:

\textbf{(1) Changing APIs.} While major data science libraries have stabilized in popularity, their APIs continue to undergo rapid iterations and frequent updates. Changes in the behavior of specific functions directly influence provenance patterns. For example, the shift from pandas version 1.x to 2.x introduced significant changes, including the removal of the `inplace=True` parameter in certain methods, affecting provenance tracking strategies. Thus, any provenance capture system must be flexible enough to adapt seamlessly to evolving APIs.

\textbf{(2) Diverse Operations.} Data science pipelines include a much broader range of operations compared to traditional database operations \cite{semirings2007}. These pipelines frequently involve diverse operations such as complex reshaping (e.g., NumPy’s `reshape` and `transpose` functions), convolution functions (e.g., pandas' `apply` method), and transformations utilizing machine learning models (e.g., predictions from scikit-learn models). Capturing provenance across these varied operations requires a more generalized and flexible provenance model.

\textbf{(3) Scale of Datasets.} The enormous size and dimensionality of contemporary datasets introduce significant complexity to provenance capture. For example, tracking cell-level provenance in large-scale genomics datasets, often containing millions of rows and thousands of columns, can be computationally expensive and storage-intensive. Therefore, efficient provenance capture mechanisms must manage storage overhead, scale effectively, and maintain performance without sacrificing accuracy.

In this paper, we discuss our experiences addressing these challenges. Specifically, we describe our implementation of a prototype cell-level provenance capture tool for the \texttt{numpy} library. This tool is part of the larger DSLog system for array provenance management \cite{dslog}. While DSLog has been previously introduced, this is the first presentation of the tool's implementation and performance. The core contribution of this paper is demonstrate that efficient memory management can significantly reduce provenance capture overhead, making real-time provenance capture feasible for larger-scale array data operations.

Firstly, we demonstrate cell-level provenance by annotating every cell with a dynamic array representing its lineage. Secondly, by sequentially allocating memory for this dynamic array and pre-allocating a memory buffer, we achieve a reduction in annotation latency of up to 275x compared to a naive implementation. Lastly, although our current implementation targets only the \texttt{numpy} library, we conclude by briefly discussing how our approach could be universally extended to other libraries and data types, along with the implications of universal provenance capture systems for data science research.

\subsection{Motivating Example}
\begin{figure}
    \centering
    \includegraphics[width=0.49\columnwidth]{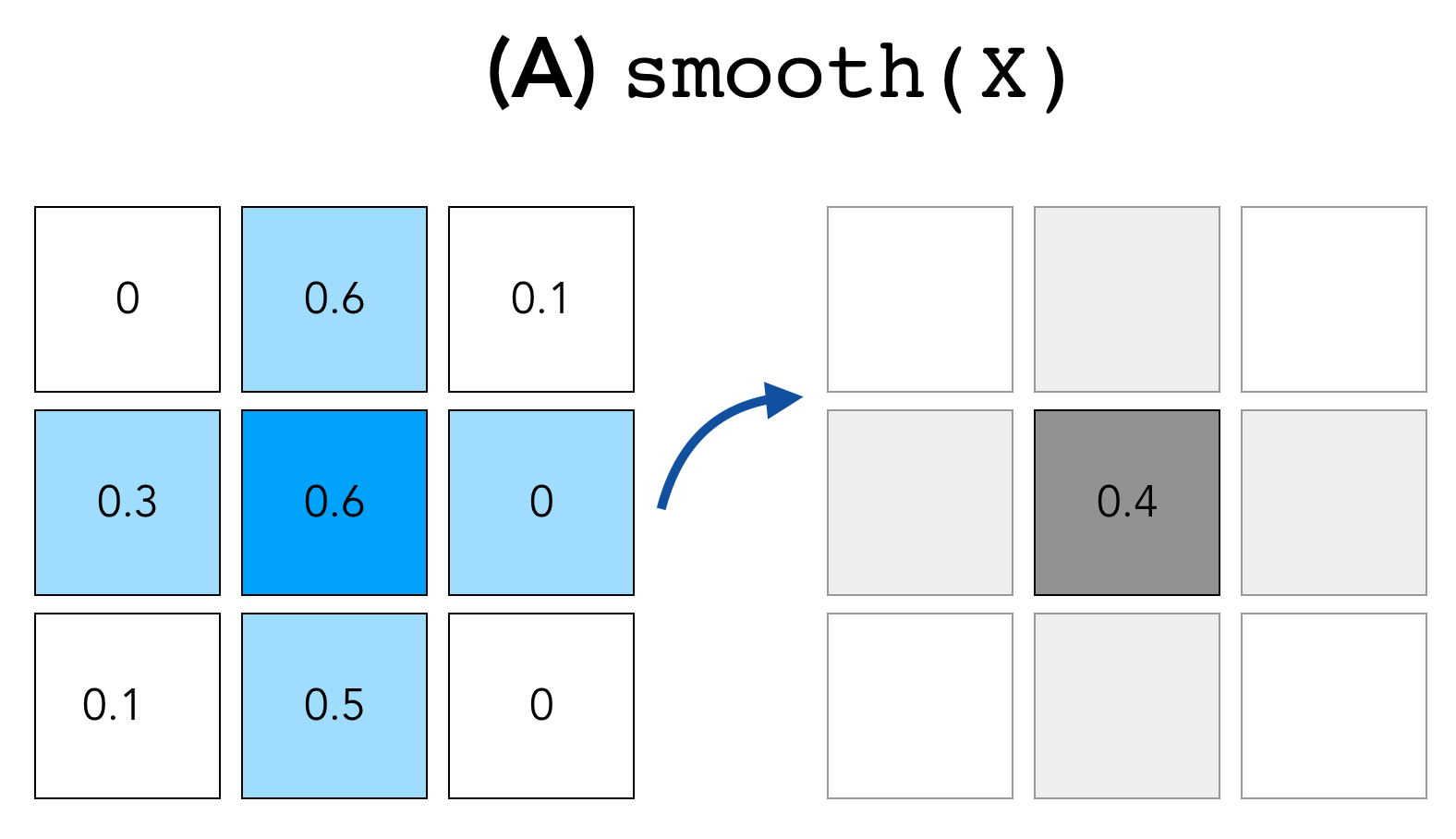}
    \includegraphics[width=0.43\columnwidth]{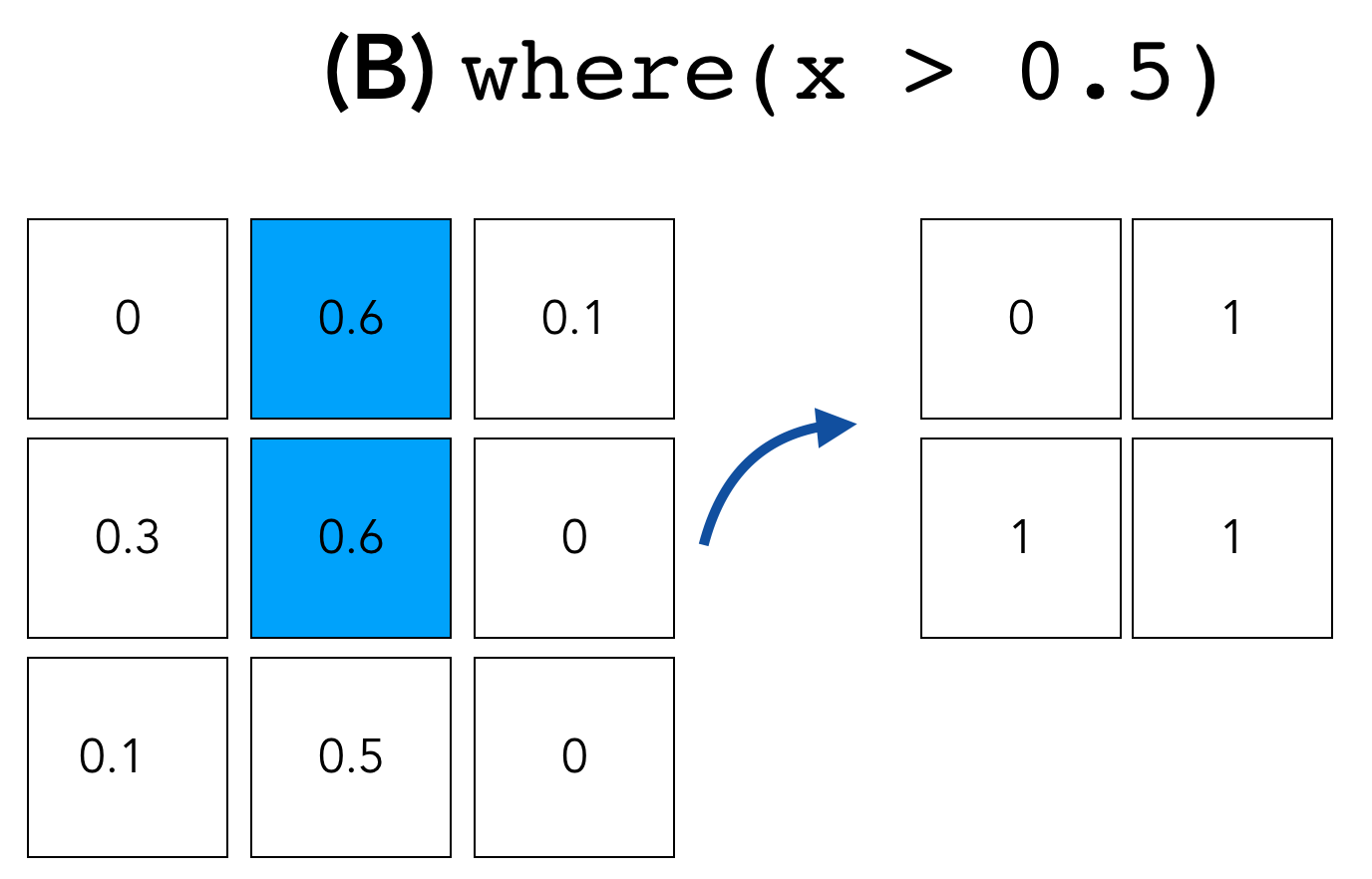}
    \caption{Visual illustration of two non-standard array operations common in imaging and scientific applications. \label{fig:example}}
\end{figure}

\begin{figure}[htbp]
\centering
\includegraphics[width=0.7\columnwidth]{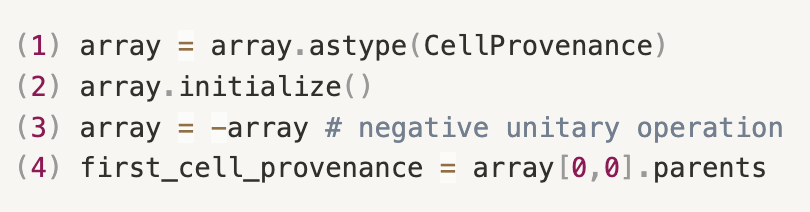}
\caption{Example of provenance capture.}
\label{fig:code_example}
\end{figure}

Consider an imaging application where a sensor outputs a 2D array representing pixel intensities ranging from 0 to 1. Such scenarios frequently occur in industrial defect detection, thermal imaging, and astronomy. A common analysis involves identifying pixels whose intensities significantly surpass those of their neighbors, signaling potential areas of interest.

This analysis can be represented through a simplified array-based pipeline:
\begin{lstlisting}
def hotspot(X):
    X_smoothed = smooth(X)
    hotspots = where(X_smoothed > 0.5)
    return hotspots
\end{lstlisting}
Here, \texttt{smooth(X)} applies a smoothing filter (e.g., calculating mean intensity within a neighborhood), and \texttt{where(condition)} generates coordinates for pixels meeting the specified condition. Figure \ref{fig:example} visually demonstrates these operations.

From a provenance perspective, users may seek to trace the origin of a detected hotspot back to the original pixels contributing to its detection. Each output pixel identified by the \texttt{where} operation depends on a neighborhood of pixels processed by \texttt{smooth}. Capturing this relationship requires recording precise cell-level provenance, linking each output pixel back to the corresponding region in the input array.

DSLog explicitly captures such cell-level provenance via bipartite mappings between input and output array cells. With these mappings, users can efficiently trace any hotspot back to its source pixels. Our prototype provenance capture system, integrated into the \texttt{numpy} library, automatically generates these mappings during pipeline execution. This prototype specifically targets capturing the union of all input cell contributions to output cells.

\section{Provenance Capture}
\subsection{Python Implementation}

We first introduce our Python implementation of annotation-based provenance capture to demonstrate the core workflow for generating tracking information.

We define a custom \texttt{CellProvenance} data type for the \texttt{numpy} library. This data type includes typical scalar data and a list of cell indices indicating dependencies from previous cells. Figure \ref{fig:code_example} shows an example of the capture workflow with CellProvenance.

\begin{lstlisting}
# annotated data type
class CellProvenance:
    data_value: Any
    parents: List[Indices]
\end{lstlisting}

Before an operation is performed, we first create an array using this data type \textbf{(1)}. Next, we initialize the \texttt{parents} property with the original array indices \textbf{(2)}; for instance, the top-left cell in the array could be annotated with the index (0, 0).

\begin{lstlisting}
# provenance initialization
c.parents = [(c.index1, c.index1)]
\end{lstlisting}

Each array operation in the \texttt{numpy} library is extended so that the \texttt{parents} property of the output value becomes the union of the \texttt{parents} properties from the input values \textbf{(3-4)}. This effectively captures lineage information for all numpy operations.

\begin{lstlisting}
# provenance operation 
c.parents = Union(a.parents, b.parents)
\end{lstlisting}

Given an annotated output array with each cell annotated with provenance, DSLog can generate a compressed relational table that represents provenance for that operation.

Although relatively straightforward, this implementation addresses two of the three challenges discussed in the introduction. Since our provenance annotations integrate easily with any \texttt{numpy} operation (achievable through simple decorator declarations in Python), it is robust to changing APIs. Moreover, since provenance is captured in real-time at the cell level, it effectively can represent diverse operations.

However, this annotation method is computationally intensive because the provenance operation is applied individually to each cell, and it scales poorly as the original array size increases. To improve this, our DSLog implementation more efficiently manages memory operations involved in annotations.

\subsection{C Implementation}

\begin{figure}[htbp]
\centering
\includegraphics[width=0.8\columnwidth]{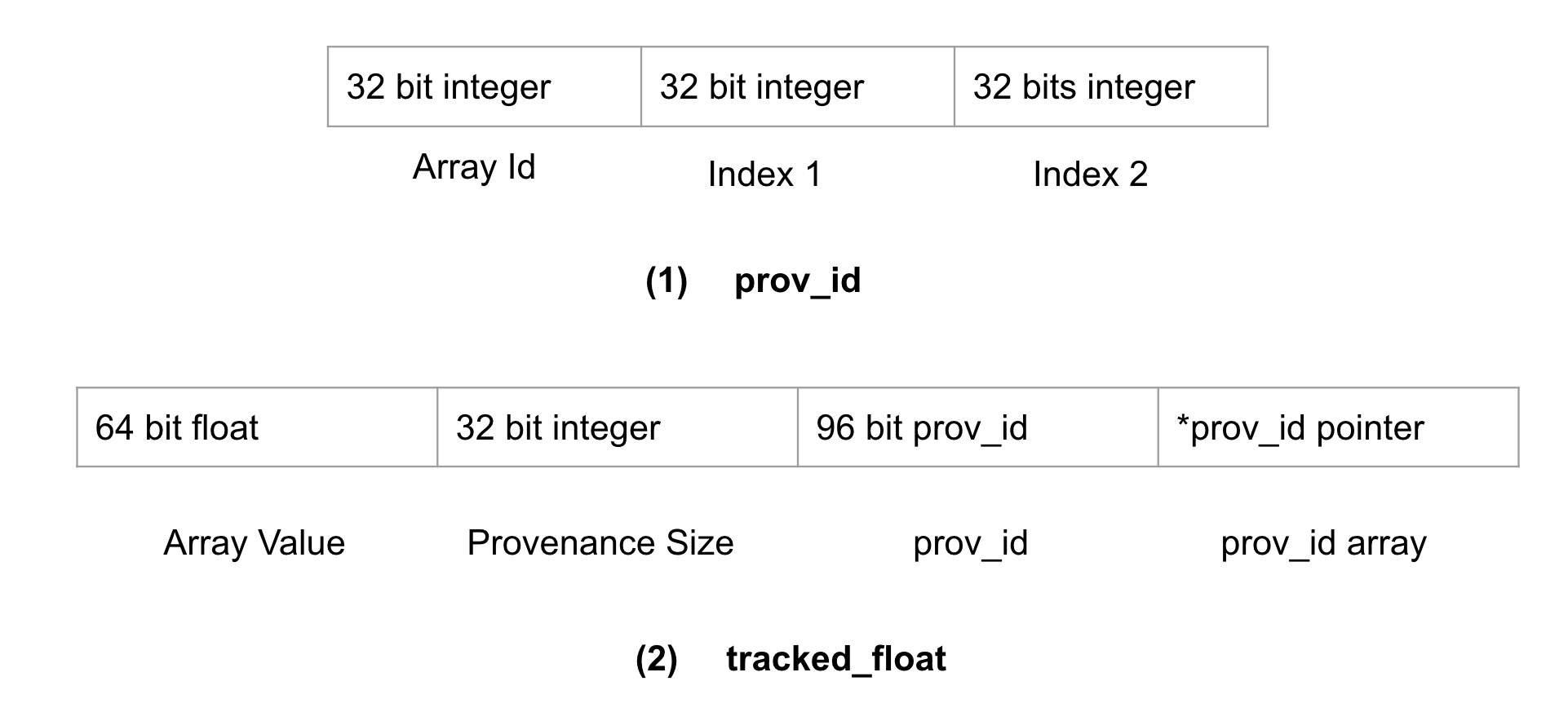}
\caption{Diagram of DSLog's tracked\_float memory structure.}
\label{exp:structure}
\end{figure}

To overcome performance limitations in annotation, we introduce the \texttt{tracked\_float} data type. The structure of this data type is shown in Figure \ref{exp:structure}. Each provenance annotation consists of three 32 bit integers: the first indicates the array ID, and the second and third represent array indices, uniquely identifying cells for arrays with up to two dimensions (though this strategy generalizes to higher dimensions). These tuples are stored in a C array, with the first annotation directly embedded in the \texttt{annotated\_cell}, and subsequent annotations stored as pointers to dynamically allocated memory buffers. In details, when the number of annotations exceed the current buffer, memory is explicitly re-allocated to accommodate a larger array. Our system allows pre-allocating memory buffers for specific functions to reduce the frequency of memory allocation requests, and we plan to implement better dynamic buffer management in the future. 

To implement provenance capture for all \texttt{numpy} operations using this data structure, we only need to adjust two primitive data operations. For unary operations, we copy the \texttt{prov\_id} and pointer values to the output. For binary operations, we concatenate the \texttt{prov\_id} arrays from both inputs. This permits duplicate annotations, but subsequent processing can de-duplicate the provenance (which is implemented in the full DSLog system).

Our preliminary experiments show that this implementation significantly reduces the performance overhead associated with provenance annotations, effectively addressing the third challenge.
\section{Microbenchmark Experiments}
In this section, we present microbenchmark experiments to evaluate the performance overhead of the \texttt{tracked\_float} annotation. We compare DSLog against two baselines in an ablation study to demonstrate the relative improvements provided by our performance optimizations. 
\begin{itemize}
\item \textbf{Python Baseline.} A Python-implemented baseline with the exact element annotation behavior as our implementation. It captures all provenance information through Python bindings and function overrides in \texttt{numpy} as described in Section 2.1.

\item \textbf{C Baseline.} This baseline uses the same implementation of \texttt{tracked\_float} but without pre-allocated memory buffers.

\end{itemize}

These experiments are performed on \texttt{numpy} arrays containing up to 100 million cells.

\subsection{Initialization Cost}

\begin{figure}[htbp]
\centering
\includegraphics[width=0.6\columnwidth]{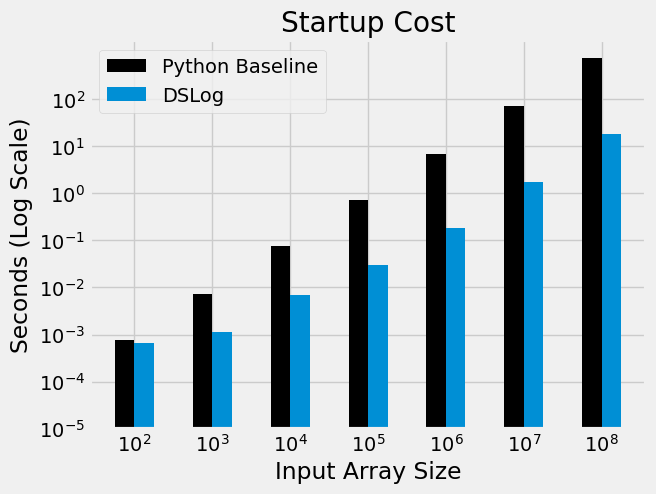}
\caption{Startup cost of provenance annotation.}
\label{exp:init}
\end{figure}

Annotated execution involves an initialization cost—the time required to convert input arrays from numerical data types into annotated data types. Figure \ref{exp:init} illustrates the annotation initialization cost for the Python baseline and DSLog (the C baseline is omitted as it behaves identically to DSLog). This cost includes converting arrays from \texttt{float} to \texttt{tracked\_float} and initializing annotations with each cell's index. The figure shows that DSLog significantly outperforms the Python baseline by approximately an order of magnitude. The Python baseline incurs additional overhead by initializing full Python objects for each cell. Conversely, DSLog directly implements a \texttt{numpy} data type stored within the array. These results indicate that Python object overhead is primarily responsible for this performance difference. Both implementations scale linearly with the number of cells, as expected, due to iteration across all cells during initialization. There is substantial potential for vectorization improvements in future work.

\subsection{Execution Cost}
\begin{figure}[htbp]
\centering
\includegraphics[width=0.6\columnwidth]{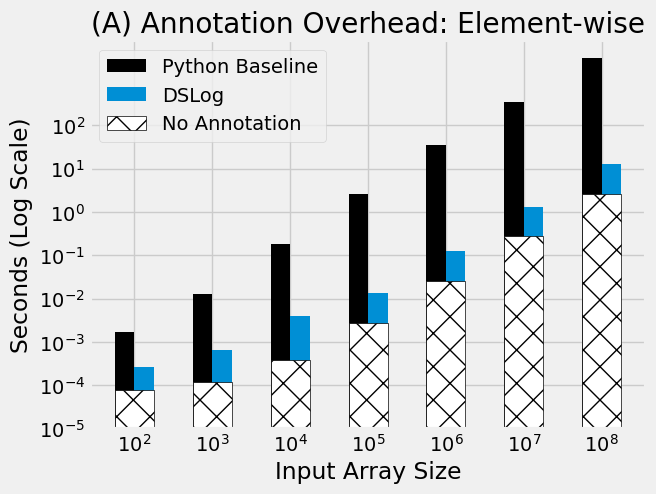}
\vspace{0.7cm} 

\centering
\includegraphics[width=0.6\columnwidth]{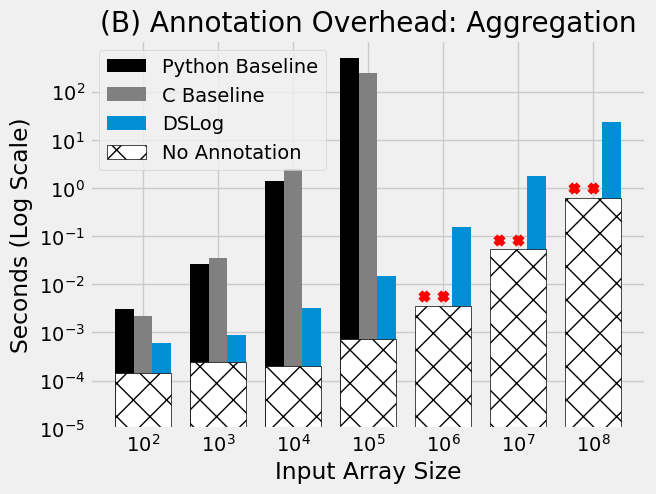}
\caption{Execution overhead of provenance annotation.}
\label{exp:annotate}
\end{figure}

We also evaluate the overhead introduced by provenance annotations during the execution of operations. This evaluation covers two extremes of provenance patterns: an element-wise function and an aggregate function. In the element-wise case, each output cell contains a single provenance annotation. In the aggregate case, each output cell aggregates annotations from input cells along an entire axis. Figure \ref{exp:annotate} shows the results.

Figure \ref{exp:annotate}(A) illustrates the overhead of the element-wise function. In this scenario, the C baseline and DSLog exhibit identical performance, so we omit the C baseline. The results demonstrate that DSLog performs up to 275x faster than the Python baseline, though it remains up to 5x slower compared to the bare function execution.

Figure \ref{exp:annotate}(B) presents the other extreme, where each output cell aggregates annotations along a whole axis. Without optimizations, the baselines fail at larger scales, including the C baseline without our optimized pre-allocated buffer. This failure occurs because the baseline continuously reallocates memory during binary operations within reduction functions. In aggregate operations, our tool is up to 34000x faster than the Python baseline but remains up to 44x slower compared to execution without annotations.

Overall, DSLog scales more effectively to larger array sizes compared to the baselines, maintaining consistent overhead for element-wise functions, while the Python baseline's overhead dramatically increases with scale.
 
\subsection{Summary of Experiments}

Although capturing provenance at the cell level inherently carries performance overhead, we have shown that careful, detailed memory management can significantly mitigate these costs compared to naive implementations. Importantly, our approach adds less than 50 seconds of overhead on aggregate operations involving roughly 100 million cells. In contexts where provenance tracking is crucial, this level of overhead could be acceptable.

\section{Future Work}
We view the current \texttt{numpy} provenance capture prototype and the broader DSLog system as steps toward a universal cell-level provenance system that is independent of specific data structures and operations. There are numerous opportunities for future research:

\textbf{Optimizing Provenance Capture.} Many additional opportunities exist to enhance the performance of provenance capture in DSLog. Provenance operations can often be parallelized, significantly reducing latency. Improved predictive memory allocation could also minimize overhead associated with dynamic reallocation of the provenance array. 

\textbf{A Universal Provenance System.} Currently, numerous data structures represent structured data, such as arrays, dataframes, relational tables, and dictionaries. Within a single data science project, data frequently transitions between these structures \cite{zhao2024quantifyingdatascienceworkflows}. Extending the ideas presented in this paper to capture provenance across different data structures would enable comprehensive end-to-end cell-level provenance tracking for all data transformations in a project.

Capturing end-to-end provenance would also open additional possibilities for data governance, raising broad research questions: (1) Does centralized provenance governance introduce privacy risks? (2) With advances in machine learning, could provenance provide regulation or proof-of-work mechanisms? (3) Can provenance help automate data science workflows and prevent statistical errors \cite{data_leakage_2012, Grafberger2021LightweightIO}?

\textbf{Partial Provenance.} For black-box operations, the simplest provenance assumption is that all inputs contribute to all outputs \cite{Wu2013SubZeroAF}. However, there may be scenarios where partial knowledge about an operation is available without precise lineage relationships between input and output cells. An extension of DSLog—which currently focuses exclusively on exact bipartite relationships—could explore partial or uncertain cell-level provenance. For example, a function signature might indicate that a filter operation occurred without specifying the exact filtered cells. Alternatively, machine learning algorithms might predict with uncertainty that an operation exhibits a one-to-one mapping by analyzing input-output patterns. We forsee future systems that effectively capture, record and query partial cell-level provenance, and are excited to explore potential applications of this information.

\section{Related Works}
There has been various previous works exploring the capture of provenance and lineage in data science and database operations. These works use different approaches that tackle a wide range of workflows and capture provenance at different levels of granularity. High-level provenance can often be expressed as directed acyclic graphs that are directly captured without additional annotation \cite{Missier2008DataLM, Heinis2008EfficientLT, Alexe2006SPIDERAS, Cheney2009ProvenanceID, Namaki2020VamsaTP, Grafberger2021LightweightIO, Pina2023DeepLP, zhao2024compressioninsituqueryprocessing, Vartak2018MISTIQUEAS}. For finer-grained annotations, many systems that allow for undefined types of operations use human-specified annotations to specifically describe the provenance patterns of particular functions \cite{ Grafberger2021LightweightIO, Namaki2020VamsaTP, Zhang2017DiagnosingML, Wu2013SubZeroAF, Chiticariu2005DBNotesAP, Missier2008DataLM, Heinis2008EfficientLT}. Similarly to our capture tool, some systems also support lineage annotations in every field, allowing detailed lineage tracking without logically defining operations \cite{data_leakage_2012, Grafberger2021LightweightIO, Interlandi2015TitianDP}. Systems that specifically support a defined set of operations, such as in relational systems,  may directly define the provenance pattern or algebra for each individual operation \cite{semirings2007, modin_2020, Chapman2020CapturingAQ, Psallidas2018SmokeFL, Zhang2017DiagnosingML, Yan2016FinegrainedPF, Cui2000PracticalLT, orchestra2008}. DSLog bridges the gap between human-specified annotations and detailed lineage annotations by automatically detecting common data science patterns in the raw lineage \cite{zhao2024compressioninsituqueryprocessing}. 

Specific optimizations have been explored to reduce the capture and storage costs of lineage. Hippo separates lineage metadata from the core dataframes in Spark to selectively capture lineage as needed for downstream applications \cite{Zhang2017DiagnosingML}. Smoke stores lineage data for relational operations as arrays or linked arrays to reduce storage and transcription costs \cite{Psallidas2018SmokeFL}. MLInspect directly inserts annotations into Pandas dataframes with lightweight monkey-patching to track row-level lineage \cite{Grafberger2021LightweightIO}. Titian optimizes in-memory management to track lineage for Spark operations \cite{Interlandi2015TitianDP} by directly leverage Spark's BlockManager for distributed memory management, and selectively writing to disk. 
\section{Conclusion}
In this paper, we discuss the implementation and performance of our prototype \texttt{numpy} provenance capture system. Our preliminary results suggest that it is possible to explicitly capture cell-level provenance in large-scale data operations with reasonable latency through efficient memory management. Since annotations are over basic \texttt{numpy} operations, fewer annotations are needed to provide coverage over the entire API. The flexibility of memory preallocation for known aggregations additionally increases the performance and scalability of the capture system. We see as this early work towards exploring the performance boundaries of a comprehensive low-maintenance real-time system that for detailed cell-level provenance.

\bibliographystyle{ACM-Reference-Format}
\bibliography{references}

\end{document}